\documentstyle[aps]{revtex}
\begin{document}
\title{Possible fractal structure of exact universal exchange-correlation
potential}

\author{Peter Babinec}

\address{Department of Biophysics and Chemical Physics, Comenius University, \\
 Mlynsk\'{a} dolina F1, 842 48 Bratislava, Slovakia.\\
E-mail address: babinec@fmph.uniba.sk\\
Fax: +421-2-65412305}

\maketitle

\begin{abstract}
{\bf Abstract:} Heuristic arguments are presented to show that the basic quantity of
density functional theory - exact universal exchange-correlation potential should have
fractal structure.
\end{abstract}
\vspace{0.8cm}
\newcommand{\be}{\begin{equation}}
\newcommand{\ee}{\end{equation}}
\newcommand{\bea}{\begin{eqnarray}}
\newcommand{\eea}{\end{eqnarray}}
\newcommand{\bi}{\bibitem}
\newcommand{\eps}{\epsilon}
\newcommand{\r}{({\bf r})}
\newcommand{\rp}{({\bf r'})}
\newcommand{\rpp}{({\bf r''})}
\newcommand{\rrp}{({\bf r},{\bf r}')}
\newcommand{\ua}{\uparrow}
\newcommand{\da}{\downarrow}
\newcommand{\la}{\langle}
\newcommand{\ra}{\rangle}
\newcommand{\s}{\sigma}

The fundamental equation describing electronic structure of matter is N-electron
Schr\"odinger's equation
\be
\left[\hat{T} + \hat{V} +\hat{U}  \right]
\Psi({\bf r}_1,{\bf r}_2 \ldots, {\bf r}_N)
=E \Psi({\bf r}_1,{\bf r}_2 \ldots, {\bf r}_N),
\label{mbse}
\ee
where the kinetic energy operator
\be
\hat{T} = -\frac{\hbar^2}{2m} \sum_i \nabla^2_i, \ee
electron-nuclei interaction
potential
\be
\hat{V}=\sum_i v({\bf r}_i)= -\sum_{ik}\frac{Z_ke^2}{|{\bf r}_i-{\bf R}_k|}, \ee the
sum on $k$ run  over all nuclei in the system, each with charge $Z_ke$ and position
${\bf R}_k$. And  for a Coulomb system we have electron-electron interaction potential
\be
\hat{U}=\sum_{i<j} U({\bf r}_i,{\bf r}_j)=
\sum_{i<j} \frac{e^2}{|{\bf r}_i-{\bf r}_j|}.
\ee
The usual quantum-mechanical approach to Schr\"odinger's equation (SE) can be
summarized by the following sequence
\be
v\r \stackrel{SE}{\Longrightarrow} \Psi({\bf r}_1,{\bf r}_2 \ldots, {\bf r}_N)
\stackrel{\la \Psi| \ldots|\Psi\ra}{\Longrightarrow}
{\rm observables},
\ee
i.e., one specifies the system by choosing $v\r$, put it into Schr\"odinger's
equation, solves that equation for the wave function $\Psi$, and then
calculates expectation values of observables with this wave function.
Among the observables that are calculated in this way is also  density
\be
n\r = N \int d^3r_2 \int d^3r_3 \ldots \int d^3r_N \Psi^*({\bf r},{\bf r}_2 \ldots,
{\bf r}_N) \Psi({\bf r},{\bf r}_2 \ldots, {\bf r}_N). \label{densdef} \ee Many
sophisticated methods for solving Schr\"odinger's equation have been developed, for
example, diagrammatic perturbation theory \cite{porucha}, coupled cluster method
\cite{cc}, or Brillouin-Wigner coupled cluster method \cite{bwcc}.

Nevertheless there is an attractive alternative to these methods, the
density-functional approach (DFT) \cite{parryang,gunjones,capelle,nobel}.

At the heart of DFT is the Hohenberg-Kohn theorem \cite{hk}. This theorem states
that for ground states Eq.~(\ref{densdef}) can be inverted: given a {\it ground-state}
density $n_0\r$ it is possible, in principle, to calculate the corresponding {\it
ground-state} wave function $\Psi_0({\bf r}_1,{\bf r}_2 \ldots, {\bf r}_N)$. This
means that $\Psi_0$ is a functional of $n_0$.

Recalling that the kinetic and interaction energies of a
nonrelativistic Coulomb system are described by universal operators, we can
 N-electron energy functional $E_v$ write as
\be
E_v[n] = T[n] + U[n] + V[n],
\ee
where $T[n]$ and $U[n]$ are {\it universal functionals}.
On the other hand,
\be
V[n] = \int d^3r\, n\r v\r
\label{vdef}
\ee
is obviously nonuniversal (it depends on $v\r$, i.e., on the system under
study), but simple: once the system is specified, i.e., $v\r$
is known, the functional $V[n]$ is known explicitly.
All one has to do is to minimize $E_v[n]$ with respect to $n\r$.
The minimizing function $n_0\r$ is the system's ground state charge density
 and the value $E_{v,0}=E_v[n_0]$ is the ground state energy.
The problem is that the
minimization of $E_{v}[n]$ is that the exact functional is not known and one
needs to used approximations for $T[n]$ and $U[n]$.

In the Kohn-Sham approach
the kinetic energy $T[n]$, is expressed in terms of the single-particle orbitals $\phi_i\r$
of a noninteracting system with density $n$, as
\be
T_s[n] = -\frac{\hbar^2}{2m} \sum_i^N \int d^3r\,
\phi_i^*\r \nabla^2 \phi_i\r,
\label{torbital}
\ee
because for noninteracting particles the total kinetic energy is just the sum
of the individual kinetic energies. Since all $\phi_i\r$ are functionals of
$n$, this expression for $T_s$ is an explicit orbital functional but an
implicit density functional, $T_s[n] = T_s[\{\phi_i[n]\}]$.

We now rewrite the exact energy functional as
\be
E[n] = T_s[n] + U[n] + V[n] = T_s[\{\phi_i[n]\}] + U_H[n] + E_{xc}[n] + V[n],
\label{excdef} \ee
where $E_{xc}[n]$, is called the {\it exchange-correlation} (xc) energy
functional, which incorporates correlations due to the Pauli exclusion
principle, Coulomb repulsion, and also correlation contribution to the kinetic
energy not included in  Eq.~(\ref{torbital}).

Since $T_s$ is now written as an orbital functional one cannot directly
minimize Eq.~(\ref{excdef}) with respect to $n$. Instead a scheme suggested by
Kohn and Sham \cite{ks},  is commonly employed for performing
the minimization indirectly. This scheme starts by writing the minimization
as
\be
0=\frac{\delta E}{\delta n} = \frac{\delta T_s}{\delta n} + \frac{\delta V}{\delta n} +
\frac{\delta U_H}{\delta n} + \frac{\delta E_{xc}}{\delta n} = \frac{\delta T_s}{\delta
n} + v\r + v_H\r + v_{xc}\r. \ee As a consequence of Eq.~(\ref{vdef}), $\delta
V/\delta n = v\r$, the `external' potential the electrons move in. The term $\delta
U_H/\delta n$ simply yields the Hartree potential $\int d^3r'\, \frac{n\rp}{|{\bf r}-{\bf r'}|}$.
The term $v_{xc}\r = \delta E_{xc}/\delta n$ is universal exchange-correlation potential.

Consider now a noninteracting system of particles moving in external
potential $v_s\r$.
For this system the minimization condition is simply
\be
0=\frac{\delta E}{\delta n_s} = \frac{\delta T_s}{\delta n_s} + v_s\r,
\ee
since there are no Hartree and $xc$ terms in the absence of interactions.
Comparing this with the previous equation we find that both minimizations
have the same solution $n_s\r \equiv n\r$ if $v_s$ is chosen to be
\be
v_s\r = v\r + v_H\r + v_{xc}\r. \label{vsdef} \ee
Consequently, one can calculate the
density of the interacting (many-body) system in potential $v\r$ by solving the
equations of a noninteracting (single-body) system in potential $v_s\r$.
In particular, the Schr\"odinger equation of this auxiliary system,
\be
\left[ -\frac{\hbar^2 \nabla^2}{2m} + v_s\r \right] \phi_i\r = \eps_i \phi_i\r,
\label{ksequation}
\ee
yields orbitals that reproduce the density $n\r$ of the original system,

\be
n\r \equiv n_s\r = \sum_i^N |\phi_i\r|^2.
\label{ncalc}
\ee
Eqs.~(\ref{vsdef}) to (\ref{ncalc}) are the celebrated Kohn-Sham
equations.

Since both $v_H$ and $v_{xc}$ depend on $n$, which depends on the $\phi_i$, which in
turn depend on $v_s$, these equations should be solved till the self-consistency.

From converged solution $n_0$,  the total ground-state energy can be calculated using
\be
E_0=\sum_i^N \eps_i
-\frac{q^2}{2}\int d^3r \int d^3r'\, \frac{n_0\r n_0\rp}{|{\bf r}-{\bf r'}|}
-\int d^3r\, v_{xc}\r n_0\r + E_{xc}[n_0].
\ee
With the exact $v_{xc}\r$ and $E_{xc}[n]$ all many-body effects are in principle
included, which would be of tremendous importance for practical computational
applications.

Let us consider the following one-dimensional Schr\"odinger equation
\be
\left[ -\frac{\hbar^2}{2m}\frac{d^2}{dx^2} + V(x) \right] \phi_i\r = \eps_i \phi_i\r,
\label{onedim} \ee If we have set of regularly ordered eigenvalues (e.g. $\eps_i = (i
+ \frac{1}{2})\hbar \omega$ for harmonic oscillator), the local potential $V(x)$ is
given as a simple continuous function ($V(x)=\frac{1}{2}m\omega x^2$ in this case).
But let us consider that the infinite set $\eps_i$ of eigenvalues is chosen randomly.
The mystery of how a one-dimensional integrable system can produce any given
eigenvalues is resolved by adopting \cite{Wu90} the concept of a fractal potential
$V(x)$ which in the limit of infinite number of eigenvalues would have dimension
higher than one. This principle can be illustrated on the example of nontrivial
Riemann zeros, which were reproduced as an eigenvalues of Schr\"odinger equation with
potential having dimension $d=1.5$ \cite{Wu93}, or on infinite the sequence of prime
numbers, where $d=1.8$ \cite{yzl}. It has been even suggested that in the limit of
infinity number of eigenvalues the fractal dimension should converge to $d=2$
\cite{ram}. In the Kohn-Sham equation exchange-correlation potential $v_{xc}\r$ should
be universal, which mean that the exact values of ground-state density and energy
should be obtained for uncountable infinity of $v\r$ (all possible numbers of various
nuclei at all possible positions in the space). Our suggestion is that in the analogy
with the previous case this would be possible only when $v_{xc}\r$ is a fractal
function.

it is possible to invert Kohn-Sham equation and obtain  exchange-correlation potential
for a given N-electron system from highly correlated electronic densities computed
with traditional methods of quantum chemistry. As has been recently found, commonly
used approximate functionals have potentials that are not even remotely close to the
potentials reconstructed from very precise {\it ab initio} calculations on the
two-electron atoms \cite{knowles}. Nevertheless it should be stressed that such an
exchange-correlation potential is constructed for a given particular system (atom) and
hence is not free from an implicit dependence on the given particular external
potential and obtained expression for $v_{xc}$ is exact only for given atom  and for
all other atoms or molecules constitutes only an approximation to the ground-state
density. In this sense such an approach is analogous to the local-density
approximation which similarly replaces exact universal exchange-correlation potential
by that of homogeneous electron gas with the same local density. It is natural that
potentials obtained in such a way are different from approximate functionals because
these are constructed in order to be universal (although approximate) for all atoms
and molecules. The exact universal potential would be also different from such a
particular atomic potentials.

For the reconstruction of genuine {\bf universal} potential 10, 50 or maybe 100 atoms
or molecules should be used simultaneously for iterative (till the Kohn-Sham equations
would not reproduce exact densities for all atoms) reconstruction of $v_{xc}\r$ as
done in the analysis of Riemann zeros potential (the simplest system for such a study
are spherically symmetric atoms, a one-dimensional systems where $v_{xc}$ is just a
function of $r$, the radial distance from nuclei). Nevertheless this is not an easy
task because very high numerical precision is needed (in the reconstruction of the
potential producing Riemann zeros or primes several hundreds of eigenvalues were
needed to reveal fractal structure of a potential). Another possibility is to
represents universal $v_{xc}\r$ as a trial fractal function with several parameters to
be fitted.

Yet another possible way is to express the density-functional exchange and correlation
potential in terms of the one-particle Green's function which in turn may be expressed
in terms of the ground-state density Kohn-Sham orbitals \cite{sham}. In principle,
this self-consistent loop yields the exact $v_{xc}\r$.

The concept of a fractal potential has been already
successfully exploited in the study of protein conformational dynamics \cite{protein},
where has been found that the fractal dimension of of the total potential energy as a
function of time is almost independent of temperature, and increases with time, more
slowly the larger the protein. Perhaps the most striking observation of this study is
the apparent universality of the fractal dimension, which depends only weakly on the
type of molecular system.

It is interesting to note that both density functional theory \cite{hk,ks} and Mandelbrot concept of
fractals \cite{mand} emerged in the same time, about 40 years ago. We hope that our
suggestion that fractals are also behind the density functional theory would be useful
in searching for the exact universal exchange-correlation potential, the {\em Holy Grail}
of many body physics.

\section*{Acknowledgements}
The author would like to thank VEGA grant 1/9179/02 for the financial support.

\end{document}